\documentclass[12pt]{scrartcl}
\usepackage{graphicx} 
\usepackage{amsmath} 
\usepackage[a4paper, margin=3.0cm]{geometry}
\usepackage{authblk}
\usepackage{float}
\usepackage{amsmath}
\usepackage{comment}
\usepackage{hyperref}

\title{Correction of STEM Distortions}

\author[1,2]{Pavel Potapov}
\author[3]{Giulio Guzzinati}

\affil[1]{IFW-Dresden, Dresden, Saxony, Germany}
\affil[2]{TEMDM, Dresden, Saxony, Germany }
\affil[3]{CEOS GmbH, Heidelberg, Baden-Württemberg, Germany}

\begin{document}

\maketitle

\noindent
\textit{The manuscript considers Scanning Transmission Electron Microscopy (STEM) images and derives transformations needed to correct various distortions occurring during scanning. These transformations form the basis for the correction algorithms implemented in the CEOS Panta Rhei and TEMDM software. The manuscript is intended as a technical reference and is meant to be published only on arXiv rather than in peer-reviewed journals.}

\section{Introduction: Source of Distortions}\label{sec:intro}

A STEM image is not a snapshot, but requires a finite time to be fully acquired, with different points being measured at different times. Therefore, all dynamical variations occurring during the acquisition affect the resulting STEM image and distort the true structure.

Those variations can be classified as:
\begin{itemize}
	\item Continuous drift caused by mechanical instability of the stage or various thermal fluxes among microscope components. This drift is most pronounced shortly after inserting a holder in the microscope and decays with time, although it never completely disappears. Within short time intervals, the drift can be approximated as a linear shift of the field of view.
	\item Irregular variations caused most probably by stray electrical and magnetic fields, electronics instability, or charging. These variations may occur in a wide frequency range, from Hz to MHz, and have an unpredictable pattern.
\end{itemize}

STEM acquisition is typically performed in a raster fashion. The electron probe moves over the fast scanning direction, acquiring a row of the image, followed by shifting the row origin along the slow scan direction, and acquiring the next row. Accordingly, the fast image direction is generally the $x$ axis of the displayed image, and the slow scanning direction is the $y$ axis. The two directions can be arbitrarily rotated via the so-called \emph{scan rotation}, but the slow scanning direction always suffers from the largest distortions.

In this manuscript, we consider continuous drift and variations ranging from milliseconds to seconds. Distortions occurring within a given fast scanned line are disregarded; this is known as the \emph{rigid rows} approximation.


\section{Sequential Frames Acquisition}

In the absence of \textit{a priori} information regarding the structure or lattice geometry, it is impossible to determine if a single image is distorted or not. Consequently, multiple frames must be acquired to retrieve and correct these distortions. The correction strategy therefore depends on the assumptions about the nature of the drift and other distortions.

\subsection{Correction for Linear Drift}\label{sec:linearstack}

Consider a linear drift, which is constant in both direction and magnitude. As the field of view shifts continuously, the average position of each subsequent frame is displaced by:
\begin{align}
	\Delta_x & = dx/dt \, (T + \Delta T) \\
	\Delta_y & = dy/dt \, (T + \Delta T)
\end{align}
where $\Delta_x$ and $\Delta_y$ are $x$- and $y$-components of the shift, $T$ is the frame acquisition time, and $\Delta T$ is the pause between acquisitions.

This shift can be measured, for instance, by cross- or phase-correlating the frames. The simplest correction is to shift the frames rigidly relative to each other according to the measured shift.  However, this  only aligns the origins of the frames, whereas the frames  stay rigid. This is equivalent to the assumption that the shift is accumulated only during the pause between subsequent acquisitions and does not occur during acquisitions themselves. Apparently, the assumption of a drift constant in time  is more appropriated.

\begin{figure}[ht!]
	\includegraphics[width=0.5\textwidth]{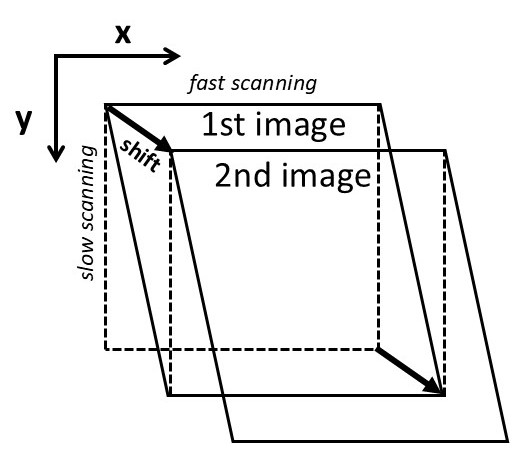}
	\caption{Sequential acquisition of two STEM images with a zero pause under conditions of constant linear drift. The true, undistorted frames are shown dashed.}
	\label{fig1}
\end{figure}

In this section, we consider  correction that involves the affine transformation of frames to cancel the \emph{linear} distortions accumulated within each individual frame. Assume for simplicity that the pause between acquisitions is absent ($\Delta T = 0$), as shown in Fig. \ref{fig1}.

Here, we use the coordinate axes common in computer vision - origin at the top-left, $x$-axis goes to the right, and $y$-axis goes downwards. Defining the components of dimensionless shift as $S_x = dx/dt \, (T/W)$ and $S_y = dy/dt \, (T/H)$, where $W$ and $H$ are the width and height of the \textit{undistorted} image in pixels, the coordinate transformation is written as:

\begin{align}
	x_1 & = x + S_x y         \\
	y_1 & = y + S_y y         \\
	x_2 & = S_x W + x + S_x y \\
	y_2 & = S_y H + y + S_y y
\end{align}
where ($x, y$) are the true (undistorted) feature coordinates in pixels and ($x_1, y_1$) and ($x_2, y_2$) are their measured coordinates in the first and second images, respectively.

If the origin of the second image is reset according to the cross-correlation ($x'_2 = x_2 - S_x W$ ; $y'_2 = y_2 - S_y H$) with the first image, its coordinates transform exactly as those in the first image:
\begin{align}
	x'_2 & = x + S_x y \\
	y'_2 & = y + S_y y
\end{align}

In matrix form, we write the transformation for any $i$-th image as
\begin{equation}
\begin{pmatrix}
    y'_i \\[4pt]
    x'_i
\end{pmatrix}
=
\begin{pmatrix}
    1+S_y & 0 \\[4pt]
    S_x   & 1
\end{pmatrix}
\begin{pmatrix}
    y_i \\[4pt]
    x_i
\end{pmatrix}
\label{eqAffineCorrection}
\end{equation}
And the inverse transformation is
\begin{equation}
	\begin{pmatrix}
		y \\[4pt]
		x
	\end{pmatrix}
	=
	\begin{pmatrix}
		1+S_y & 0 \\[4pt]
		S_x   & 1
	\end{pmatrix}^{-1}
	\begin{pmatrix}
		y'_i \\[4pt]
		x'_i
	\end{pmatrix}
\end{equation}

Therefore, the affine transformation requires only two parameters, deduced from cross-correlation, namely the $x$- and $y$-components of the shift between two subsequent frames.

\begin{figure}[ht!]
	\caption{Inverse transformation for the distortion in Fig. \ref{fig1}. The observed rectangular frame (dashed) deforms to the dark parallelogram. The largest (blue) or smallest (red) frame can be chosen for displaying the final result. Various combinations of the shifts signs are shown on the right.}
	\includegraphics[width=0.9\textwidth]{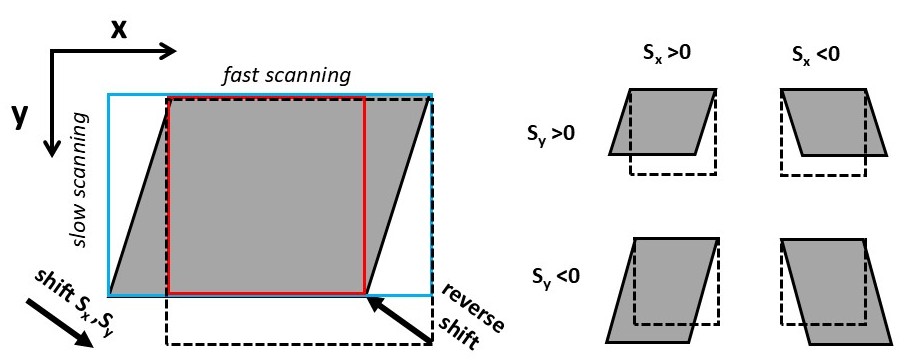}
	\label{fig_2}
\end{figure}

Fig. \ref{fig_2} shows the inverse transformation, restoring the undistorted image, using an affine transformation with sub-pixel bilinear interpolation. Such deformations for any combinations of the signs of shifts ($S_y, S_x$) are displayed as small icons at the right of the figure.

The matrix transformation above assumes constant values $S_x$ and $S_y$ among frames. At the next level of consideration, constant drift may be assumed only between two subsequent frame acquisitions. Then, the drift values must be replaced by $S^{(i)}_x$ and $S^{(i)}_y$, and the shift must be evaluated for each $i$-th and $(i+1)$-th frames pair.

\subsubsection{Deduce Linear Drift from FFT}
\label{sec:FFT_stack}

For atomically resolved images, cross-correlation might deliver ambiguous results as shifts on one or several lattice periods are sometimes indistinguishable. A more robust alternative is the analysis of their FFT transforms; however, this requires comparison with an undistorted reference. Indeed, under the assumption of constant drift, the FFT transforms of all frames are distorted in the same way and do not provide information about the true lattice geometry.

The reference must be a STEM image of the  \textit{same} lattice taken either with very fast scanning or in the situation when the drift has largely vanished. Note that this reference not strictly need to be of the same size as the multi-frame stack or represent exactly the same investigated area. In many cases, the only requirement is that the reference is taken from a lattice of the exactly same geometry.

A good example of such a reference is the survey image taken before the acquisition of a STEM EELS spectrum-image. The survey is typically taken with a much shorter dwell time than the STEM EELS dataset, thus its distortions should be negligible.

A linear drift can be deduced by comparing undistorted and distorted FFTs as follows. In a 2D power spectrum (magnitude of FFT) of an undistorted image, a given node $P$ of lattice periodicity   manifests itself  as a spot with a horizontal coordinate $u$ and a vertical coordinate $v$: 
\begin{equation}
	P = \cos(2\pi (u x + v y))
\end{equation}

In the power spectrum of a distorted image, this node shows up as a spot ($u_i, v_i$):
\begin{equation}
	P = \cos(2\pi (u_i x_i + v_i  y_i)) = \cos(2\pi (u_i(x + S_x y) + v_i(1+S_y) y))
\end{equation}

and, after rearranging the terms, we obtain:
\begin{align}
	u & = u_i                   \\
	v & = (1+S_y) v_i + S_x u_i
\end{align}
or in matrix form:
\begin{equation}
	\begin{pmatrix}
		v - v_i \\[4pt]
		u - u_i
	\end{pmatrix}
	=
	\begin{pmatrix}
		v_i & u_i \\[4pt]
		0   & 0
	\end{pmatrix}
	\begin{pmatrix}
		S_y \\[4pt]
		S_x
	\end{pmatrix}
\end{equation}

The $u$-position of the {FFT} spot does not change, thus at least two spots with differing $v$-coordinates are required for determination of  $S_x$ and $S_y$:

\begin{equation}
	\begin{pmatrix}
		v - v_{i1} \\[4pt]
		v - v_{i2}
	\end{pmatrix}
	=
	\begin{pmatrix}
		v_{i1} & u_{i1} \\[4pt]
		v_{i2} & u_{i2}
	\end{pmatrix}
	\begin{pmatrix}
		S_y \\[4pt]
		S_x
	\end{pmatrix}
\end{equation}

This system has a solution if the determinant of the coefficient matrix is non-zero:
\begin{equation}
	u_{i2} v_{i1} - u_{i1} v_{i2} \neq 0 \quad \textit{(i.e.)} \quad \frac{u_{i2}}{u_{i1}} \neq \frac{v_{i2}}{v_{i1}}
\end{equation}

In other words, the chosen FFT spots should not be higher-order copies of each other.

The precision of the shift evaluation can be increased by considering more than two FFT spots and solving the overdetermined system of equations:
\begin{equation}
	\begin{pmatrix}
		v_1 - v_{i1} \\[4pt]
		v_2 - v_{i2} \\[4pt]
		v_3 - v_{i3} \\[4pt]
		...          \\[4pt]
	\end{pmatrix}
	=
	\begin{pmatrix}
		v_{i1} & u_{i1} \\[4pt]
		v_{i2} & u_{i2} \\[4pt]
		v_{i3} & u_{i3} \\[4pt]
		...    & ...
	\end{pmatrix}
	\begin{pmatrix}
		S_y \\[4pt]
		S_x
	\end{pmatrix}
\end{equation}

The values $S_x$ and $S_y$ can be retrieved from this system in the sense of best least-squares difference.

\subsubsection{Frame Resizing and Aligning}\label{sec:linear_stack_realignment}

\begin{figure}[ht!]
	\caption{Frames shift during consequent acquisitions where (a), (b) a pure $x$-drift or (c), (d) a pure $y$-drift are present. The dashed rectangles show the nominal scanning areas while gray parallelograms show how they should be deformed to restore the true structure. The arrows show the accumulated drift among frame acquisitions. The orange area includes every pixel in all frames while the green area outlines the common rectangular area without blanked fields.}
	\includegraphics[width=\textwidth]{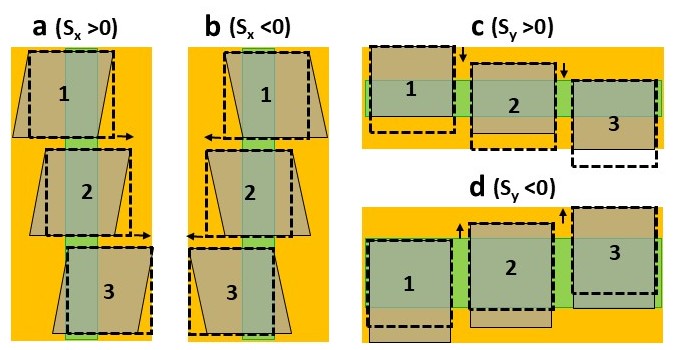}
	\label{fig94a}
\end{figure}

The accurate alignment of the corrected frames is crucial for further analysis. The inverse transformation may shift some pixels to negative $(x,y)$ coordinates or to coordinates exceeding the original image size. Such pixels are clipped during typical image processing, which keeps the frame size unchanged. It is however worth adding flexibility in order to enlarge the frame size and preserve the wider field of view. Alternatively, blank segments at the corners of an image might be undesired. Then the frame size should be reduced relative the initial one.

Fig. \ref{fig94a} shows the shift of the scanned area during the multiframe acquisitions for the simplified cases when a pure $x$ of a pure $y$ drift are present. The following rules for frame resizing can be derived from this figure:

To keep the whole deformed area in view, the size should be enlarged to ($W^L, H^L$) as:
\begin{align}
	W^L= & W (1 + \left|\sum_{i=0}^{n}{S^{(i)}_x}\right| ) & H^L & = H (1 + \left|\sum_{i=1}^{n}{S^{(i)}_y}\right| - S^{(n)}_y ) & (S_y>0) \\
	     &                                                 & H^L & = H (1 + \left|\sum_{i=1}^{n}{S^{(i)}_y} \right|- S^{(0)}_y ) & (S_y<0)
\end{align}

Correspondingly, the frame origin must be adjusted with the offsets:
\begin{align}
	x & = x_i  + W(\sum_{j=1}^{i-1}{S^{(j)}_x} -S^{(i)}_x) & (S_x>0) &   & y = y_i + H\sum_{j=0}^{i-1}{S^{(j)}_y} & \\
	x & = x_i  + W(\sum_{j=1}^{i-1}{S^{(j)}_x} )           & (S_x<0) &                                              \\
\end{align}

To exclude the blanked segments, the size should be reduced to ($W^S, H^S$) as:
\begin{align}
	W^L & = W (1 - |\sum_{i=0}^{n}{S^{(i)}_x}|)
	    & H^L                                   & = H (1 - |\sum_{i=1}^{n}{S^{(i)}_y}| - S^{(n)}_y) & (S_y>0)                                                    \\
	    &                                       & H^L                                               & = H (1 - |\sum_{i=1}^{n}{S^{(i)}_y}|- S^{(0)}_y) & (S_y<0)
\end{align}

\begin{figure}[ht!]
	\caption{Left: Synthetic square lattice sequentially scanned under a shift $S_x = 0.1 \pm 0.02$ in the $x$-direction (the shift direction denoted by yellow arrows). Right: The reverse affine transformation (green arrows show the direction of the reverse shift). The frames are enlarged to $(H^L, W^L)$ in order to accommodate the shifts and inclined corners. The common area of size $(H^S, W^S)$ containing no blanked pixels is outlined in red.}
	\includegraphics[width=0.7\textwidth]{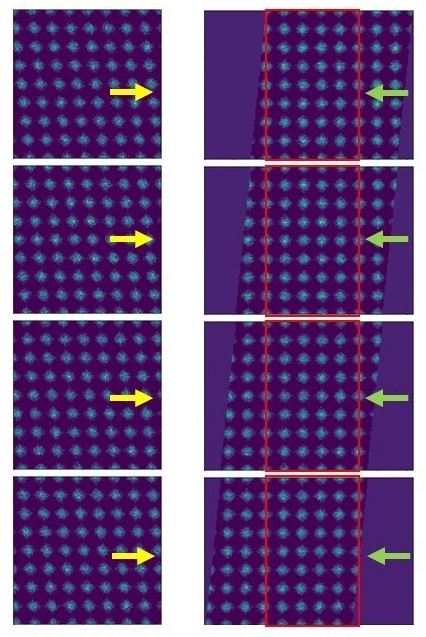}
	\label{fig94}
\end{figure}

with the offsets:
\begin{align}
	x & = x_i  + W(\sum_{j=1}^{i-1}{S^{(j)}_x} )            & (S_x>0) &   & y = y_i + H\sum_{j=0}^{i-1}{S^{(j)}_y} & \\
	x & = x_i  + W(\sum_{j=1}^{i-1}{S^{(j)}_x} -S^{(i)}_x ) & (S_x<0) &                                              \\
\end{align}

Figure \ref{fig94} demonstrates that $(W^L, H^L)$ and $(W^S, H^S)$ can noticeably deviate from the original frame size when a number of frames are accumulated. In the situation of significant shifts, the $(W^S, H^S)$ area may even disappear completely.

These resizing and offsets provide only an approximate alignment of the corrected frames. As will be shown below, the linear correction might be insufficient for atomic resolution images. Therefore, we warn against summing frames within the $(W^S, H^S)$ size. Integration is only recommended after correction for irregular, non-linear variations.

\subsection {Correction for Non-Linear Irregular Variations}\label{sec:nonlinearstack}

Suppose that a linear drift among the sequentially collected frames is negligible or accurately corrected. Still, irregular variations within each frame might distort structures in the field of view. Such distortions are typically visible only in  atomically resolved STEM images, but there, they might lead to serious misinterpretation of the lattice geometry and should be corrected by evaluation of many acquired frames as pioneered by Lewys Jones \cite{Jones2015}.

\begin{figure}[ht!]
	\caption{ Illustration of distortions in two sequentially acquired frames where stochastic variation of the origins of rigid rows is apparent. The position of any row can be compared with the position of the corresponding row in the next image and the mutual shift of the origins can be retrieved.}
	\includegraphics[width=1.0\textwidth]{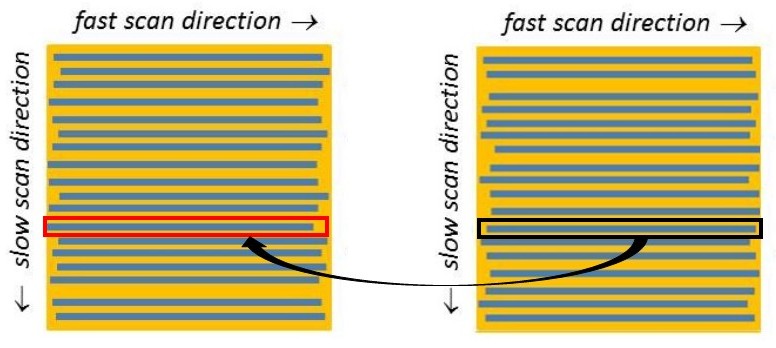}
	\label{fig5}
\end{figure}

As mentioned in Introduction, we follow the approximation of \textit{rigid rows}, as scanning of one given row is fast. However, the collection of many rows takes a much longer time, and the origins of the rows may be affected by irregular variations. This is shown schematically in Fig. \ref{fig5}.

The origins of the rows must be realigned in order to uncover the true structure. One way to do that is to correlate the origin of each row with that of a corresponding row in another frames. Let's count the row origins $s_{x}^{i}(y)$ and $s_{y}^{i}(y)$ relative the origin of the entire i-th frame and denote them  as a function of the coordinate $y$. Presume that the  frames' origins are well aligned to each other, via for instance, cross-correlation and linear drift correction.  Then, the rows origins  deviate from their true positions as:
\begin{align}
	s_{x}^{i} (y) & = s_x^*(y) + \Delta  s_{x}^{\hat{\imath}}(y) \\
	s_{y}^{i} (y) & = s_y^*(y) + \Delta  s_{y}^{\hat{\imath}}(y)
\end{align}
where $s_x^*(y)$, $s_y^*(y)$ are the true rows origins and $\Delta  s_{x}^{\hat{\imath}}(y)$, $\Delta  s_{y}^{\hat{\imath}}(y)$ are  deviations to be found for each i-th frame.

If the \textit{frames origins} are well aligned, the  deviations of \textit{row origins} $\Delta s_{x}^{i}(y)$, $\Delta s_{y}^{i}(y)$ averaged over all frames should approach zero. Therefore, the comparison of the observed origin positions in a given i-th frame with all other j-th frames indicates their deviations from the truth and, therefore allow to retrieve the true positions from (29), (30):
\begin{align}
	 & \frac{1}{n} \sum_{j=0}^{n} (s_{x}^{i}(y) - s_{x}^{j}(y)) &=\Delta  s_{x}^{\hat{\imath}}(y) - \frac{1}{n} \sum_{j=0}^{n} \Delta  s_{x}^{j}(y) &  & \approx \Delta  s_{x}^{\hat{\imath}}(y) \\
	 & \frac{1}{n} \sum_{j=0}^{n} (s_{y}^{i}(y) - s_{y}^{j}(y)) &=\Delta  s_{y}^{\hat{\imath}}(y) - \frac{1}{n} \sum_{j=0}^{n} \Delta  s_{y}^{j}(y) &  & \approx \Delta  s_{y}^{\hat{\imath}}(y)
\end{align}

This evaluation requires a significant number of frames to average,  at least 10 in practice.

\begin{figure}[ht!]
	\caption{ Frames are divided on horizontal stripes according their laticce periodicity. The cross-correlation with stripes in other frames provides the shifts values, which are assigned to the stripes centers. Interpolation over entire rows range is performed by cubic splines.}
	\includegraphics[width=0.9\textwidth]{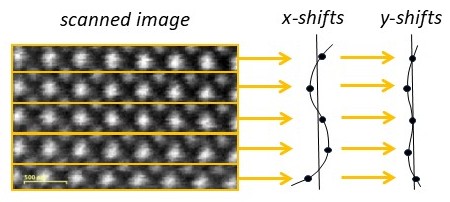}
	\label{fig6}
\end{figure}

Another question is how to correlate so many row origins among different frames? The treatment of every row is computationally expensive and, additionally, some rows might not cross the atomic positions and therefore deliver only information about noisy background.

An efficient strategy is dividing the frames on horizontal stripes according their minimal periodicity in the vertical direction as illustrated in Fig. \ref{fig6}. Such stripes can be easily cross-correlated with those in other frames providing the \textit{middle} shift of the entire stripe. The discrete $s_x(y_k)$,  $s_y(y_k)$ values serve as a skeleton for interpolation over the whole y-range. The smooth, natural interpolation can be then performed by cubic splines as sketched in Fig. \ref{fig6}.

As soon as $\Delta s_x(y)$ and $\Delta s_y(y)$ are retrieved, the true shape of images can be restored by the reverse \textit{homeomorphic} transformation. This implies a continuous deformation  with polynomial sub-pixels interpolation without tearing. We actually used the 1st order of polynomial, i.e. bilinear interpolation.

\begin{figure}[ht!]
	\caption{Fragment of a multi-frame STEM dataset corrected for in-frame non-linear variations. The linear correction was not applied in this case, but only rigid registration.}
	\includegraphics[width=0.7\textwidth]{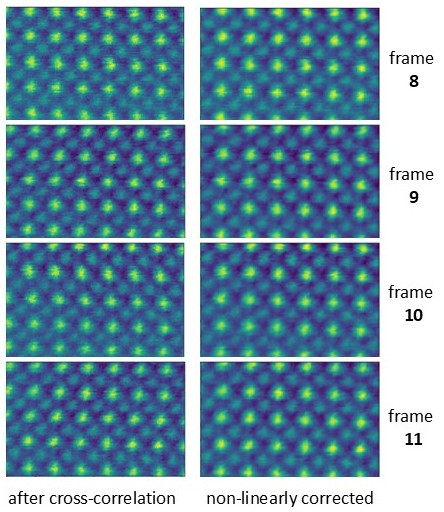}
	\label{fig.7}
\end{figure}

Such step of correction deforms the image towards the reference, however cross-correlation provides only approximate $\Delta s_x(y)$, $\Delta s_y(y)$ values, therefore under- or overcorrection is possible.  Typically, several  iterations with the gradually increasing correction power are required for the accurate reconstruction.
Fig. \ref{fig.7} demonstrates that the non-linear correction results in almost perfect alignment of the in-frame structures.

\subsubsection{Resizing and aligning frames}\label{sec:align_stack}

Similar to the case of linear correction, the back mapped area can extend over the original sizes. On the other hand, additional blank areas appear, reducing the size of rectangular frames with strictly physical image.

The updated sizes are evaluated as
\begin{align}
	W^L & = \min_i([\max(x)]_i) - \max_i([\min(x)]_i)) & \\
	H^L & = \min_i([\max(y)]_i) - \max_i([\min(y)]_i)) & \\
	W^S & = \min_i([\max(x)]_i) - \max_i([\min(x)]_i)) & \\
	H^S & = \min_i([\max(y)]_i) - \max_i([\min(y)]_i)) &
\end{align}
where $[\max(x/y)]_i$, $[\min(x/y)]_i$ are maximal or minimal calculated 'true' coordinates in the \textit{i-}-th frame.

The non-linear correction is typically accurate enough to allow summing up all available frames in order to increase the signal-to-noise ratio and accuracy of the structure evaluation. Such summation should be applied to the smallest size frames ($W^S, H^S$) with no blank segments.

\section{Alternating Scan Directions}

Xiahan Sang highlighted \cite{Sang2014} that taking several STEM images with revolving the direction of fast scan  can offer significant benefits as the directions of maximal and minimal distortions are altered.

Colin Ophus \cite{Ophus2016} realized that two images with mutually perpendicular fast scan directions are sufficient to reconstruct the ground truth. Again, the reconstruction assumes that each scan row is rigid, but the origins of rows may vary in both the x- and y- directions.

One limitation of this approach is the requirement to keep the aspect ratio of an image close to 1; otherwise the information along one of the scanning axis might appear insufficient to reconstruct the ground truth.

\begin{figure}[ht!]
	\caption{Acquisition of two images with rotating the fast scan direction 90° clockwise in between. The second image should be rotated back before treatment.}
	\includegraphics[width=0.8\textwidth]{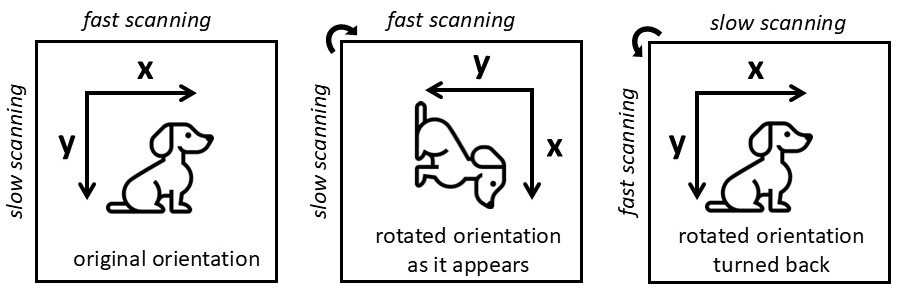}
	\label{fig.8}
\end{figure}

\subsection{Correction for Linear Drift}\label{sec:linearalternated}

Consider two subsequently acquired frames when the direction of fast scanning is rotated 90° clockwise between the acquisitions as in Fig. \ref{fig.8}. All features in the second image  appear rotated but, after digital back rotation (Fig. \ref{fig.8}, right most), it looks similar to the first image apart from the different distortions and the slight shift due to  drift.

\begin{figure}[ht!]
	\caption{ Distortions of STEM images by the constant linear drift when one image is taken in the original orientation (left) and another in the 90° clock-wise rotated orientation (right). The true, undistorted frames are shown dashed.}
	\includegraphics[width=0.8\textwidth]{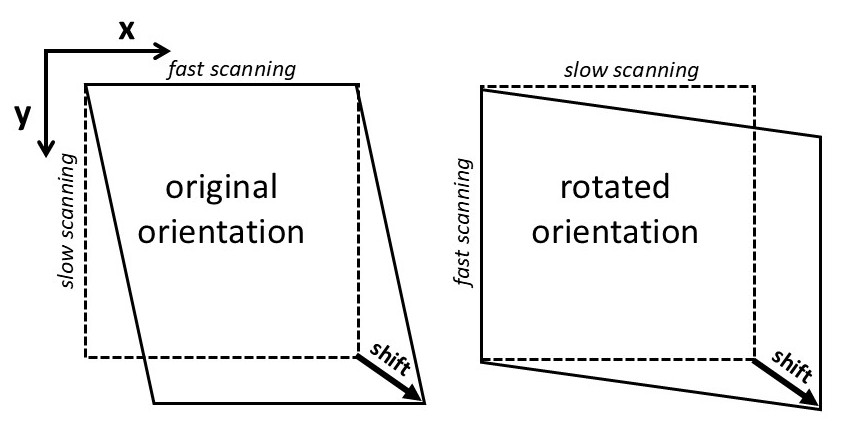}
	\label{fig_9}
\end{figure}

The transformation of coordinates in the original ($x_O, y_O$) and rotated ($x_R, y_R$) images, as guided by  Fig. \ref{fig_9}, is:

\begin{align}
	x_O & = x + S_x y         \\
	y_O & = y + S_y y         \\
	x_R & = S_x W + x + S_x x \\
	y_R & = S_y H + y + S_y x
\end{align}
where, as previously, we assume the negligible pause between  acquisitions.

Although the shapes of distorted images are different for two scan orientations, it can be shown that their geometrical centers coincide, thus cross-correlation  delivers a reasonable estimate for the linear shift between  acquired frames.

Subtracting the frame origin and writing in the matrix form:
\begin{equation}
	\begin{pmatrix}
		y_O \\[4pt]
		x_O
	\end{pmatrix}
	=
	\begin{pmatrix}
		1+S_y & 0 \\[4pt]
		S_x   & 1
	\end{pmatrix}
	\begin{pmatrix}
		y \\[4pt]
		x
	\end{pmatrix}
	\qquad
	\begin{pmatrix}
		y'_R \\[4pt]
		x'_R
	\end{pmatrix}
	=
	\begin{pmatrix}
		1 & S_y     \\[4pt]
		0 & 1 + S_x
	\end{pmatrix}
	\begin{pmatrix}
		y \\[4pt]
		x
	\end{pmatrix}
\end{equation}

The reverse transformations are:
\begin{equation}
	\begin{pmatrix}
		y \\[4pt]
		x
	\end{pmatrix}
	=
	\begin{pmatrix}
		1+S_y & 0 \\[4pt]
		S_x   & 1
	\end{pmatrix}^{-1}
	\begin{pmatrix}
		y_O \\[4pt]
		x_O
	\end{pmatrix}
	\qquad
	\begin{pmatrix}
		y \\[4pt]
		x
	\end{pmatrix}
	=
	\begin{pmatrix}
		1 & S_y     \\[4pt]
		0 & 1 + S_x
	\end{pmatrix}^{-1}
	\begin{pmatrix}
		y'_R \\[4pt]
		x'_R
	\end{pmatrix}
\end{equation}

\begin{figure}[ht!]
	\centering
	\caption{Reverse transformation for the \textit{rotated orientation} image in Fig. \ref{fig.8}, right. The observed quadratic frame (dashed) deforms to the dark parallelogram. The largest (blue) or smallest (red) frame can be chosen for displaying the final result. Various combinations of the shift signs are shown at right.}
	\includegraphics[width=0.9\textwidth]{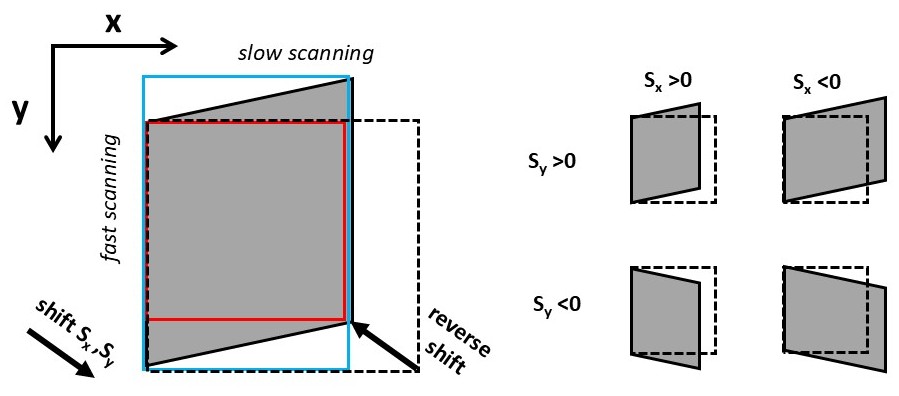}
	\label{fig.10}
\end{figure}

The reverse transformation for the original orientation image is the same as in Fig. \ref{fig_2}. In the case of the rotated image, the reverse transformation is schematically shown in Fig. \ref{fig.10}.

\subsubsection{Deduce Linear Drift from FFT}\label{sec:fftdrift}
\label{sec:FFT_rotated}

In case of atomically resolved images, cross-correlation might deliver ambiguous results as shifts on one or several lattice periods are sometimes undistinguishable. A better alternative is comparison of their Fourier transforms, that allows to retrieve the shift unambiguously in the case of alternating fast scan directions.

\begin{figure}[ht!]
	\caption{Frames acquired in the original (left) and clock-wise rotated (right) scan orientation. For better comparison, the right image is then rotated digitally back to fit the orientation of the left one. The FFTs of images show the shear of the rectangular geometry in different directions.}
	\includegraphics[width=1.0\textwidth]{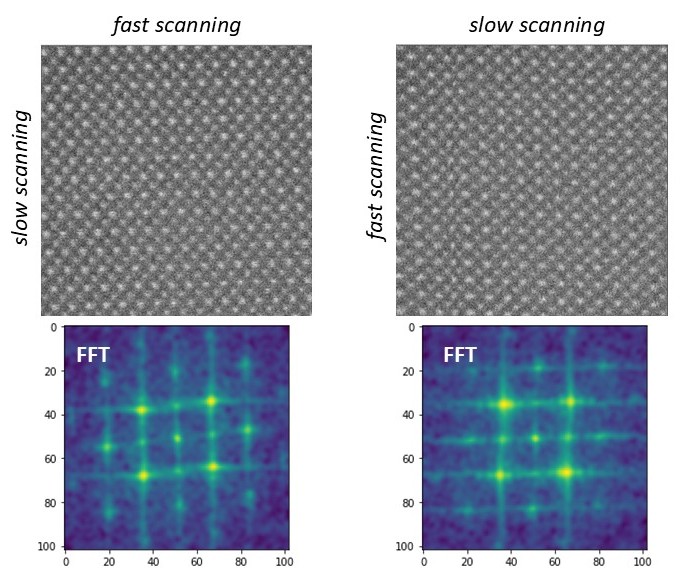}
	\label{fig_10}
\end{figure}

Fig. \ref{fig_10} shows an example of $SrTiO_3$ lattice acquired with the 90° rotated fast scans. The crystallographic planes are inclined relative to each other in two images and these distortions are clear in their FFT transforms.

In the power spectra of the original and rotated images, the periodicity node (9) manifests itself as:
\begin{align}
	P & = cos(2\pi (u_O(x + S_x y) +v_O(1+S_y) y ) = \\
	  & = cos(2\pi (u_R(1 + S_x) x +v_R(y + S_y x))
\end{align}

here, we used the coordinate transformations (41).

Rearrangement the terms results in the following system of equations:
\begin{align}
	u_O & = u_R + S_x u_R + S_y v_R \\
	v_R & = v_O + S_y v_O + S_x u_O
\end{align}
or in the matrix form:
\begin{equation}
	\begin{pmatrix}
		u_O - u_R \\[4pt]
		v_R - v_O
	\end{pmatrix}
	=
	\begin{pmatrix}
		v_R & u_R \\[4pt]
		v_O & u_O
	\end{pmatrix}
	\begin{pmatrix}
		S_y \\[4pt]
		S_x
	\end{pmatrix}
\end{equation}

The system can be solved provided that determinant of the coefficients matrix is not zero:
\begin{equation}
	u_R v_O -u_O v_R \neq 0
\end{equation}

This actually cannot be guaranteed. If the ($u,v$) spot is oriented nearly perpendicular to the shift direction, the difference between the spot positions in the original and rotated frames would be closed to zero and the solution might diverge. Therefore, it is worth to consider at least two spots ($u_1,v_1$) and ($u_2,v_2$) inserting them into the overdetermined system of equations:

\begin{equation}
	\begin{pmatrix}
		u_{O1} - u_{R1} \\[4pt]
		v_{R1} - v_{O1} \\[4pt]
		u_{O2} - u_{R2} \\[4pt]
		v_{R2} - v_{O2} \\[4pt]
	\end{pmatrix}
	=
	\begin{pmatrix}
		v_{R1} & u_{R1} \\[4pt]
		v_{O1} & u_{O1} \\[4pt]
		v_{R2} & u_{R2} \\[4pt]
		v_{O2} & u_{O2}
	\end{pmatrix}
	\begin{pmatrix}
		S_y \\[4pt]
		S_x
	\end{pmatrix}
\end{equation}

\begin{figure}[ht!]
	\caption{Images from Fig. \ref{fig_10} corrected for linear drift.}
	\includegraphics[width=1.0\textwidth]{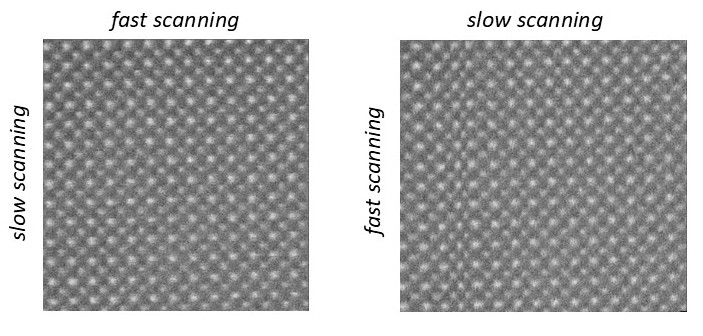}
	\label{fig_11}
\end{figure}

The values $S_x$ and $S_y$ can be retrieved   from this system in the sense of best least squared difference. With such method, the lattices shown in Fig. \ref{fig_10} can be corrected for linear drift (Fig. \ref{fig_11}).

\begin{figure}[ht!]
	\centering
	\caption{Frames shift with rotated acquisitions where (a), (b) a pure $x$-drift or (c), (d) a pure $y$-drift are present. The dashed rectangles show the nominal scanning areas while gray parallelograms show how they should be deformed to restore the true structure. The arrows show the accumulated drift among frame acquisitions. The orange area includes every pixel in all frames while the green area outlines the common rectangular area without blank fields.}
	\includegraphics[width=0.8\textwidth]{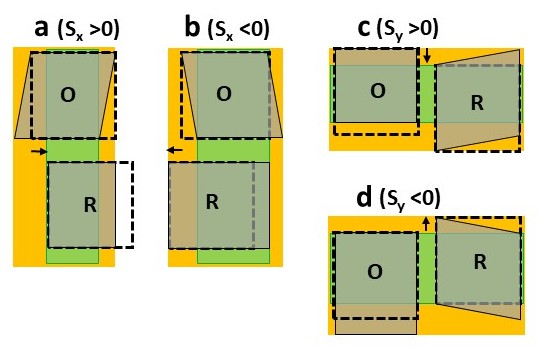}
	\label{fig_11b}
\end{figure}

\subsubsection{Frame Resizing and Alignment}

Similar to that in Section \ref{sec:linear_stack_realignment}, the frames should be resized in order to keep the whole deformed area in view or to exclude blank segments. The size common for the original and  rotated orientations the corresponding rules are derived from Fig. \ref{fig_2} and \ref{fig.10}. For simplicity, we assume a quadratic image, i.e. $W=H$ in the analysis below.

To keep the whole deformed area in view in the both original and rotated images, the size should be enlarged as
\begin{align}
	W^L & = W (1+|S_x|)  & (S_x>0) &  & H^L= W(1+|S_y|)  &  & (S_y>0) \\
	W^L & = W (1+2|S_x|) & (S_x<0) &  & H^L= W(1+2|S_y|) &  & (S_y<0)
\end{align}

with the offsets:
\begin{align}
	x & = x_O              & (S_x>0) &  &  & y = y_O          &  & (S_y>0) \\
	x & = x_R  + 2 W |S_x| & (S_x>0) &  &  & y = y_R          &  & (S_y>0) \\
	x & = x_O  + W|S_x|    & (S_x<0) &  &  & y = y_O + W|S_y| &  & (S_y<0) \\
	x & = x_R              & (S_x<0) &  &  & y = y_R          &  & (S_y<0)
\end{align}

To avoid the blanked areas, the frame size should  be reduced to:
\begin{align}
	W^L & = W (1-2|S_x|) & (S_x>0) &  & H^L= W(1-2|S_y|) &  & (S_y>0) \\
	W^L & = W (1-|S_x|)  & (S_x<0) &  & H^L= W(1-|S_y|)  &  & (S_y<0)
\end{align}

The offsets should be aligned as:
\begin{align}
	x & = x_O - 2 W S_x   & (S_x>0) &  &  & y = y_O -W|S_y|  &  & (S_y>0) \\
	x & = x_R             & (S_x>0) &  &  & y = y_R  -W|S_y| &  & (S_y>0) \\
	x & = x_O  -  W|S_x|  & (S_x<0) &  &  & y = y_O          &  & (S_y<0) \\
	x & = x_R  - 2 W|S_x| & (S_x<0) &  &  & y = y_R - W|S_y| &  & (S_y<0)
\end{align}

\subsection{Correction for Irregular Non-linear Variations}\label{sec:nonlinearrotated}
Then, the irregular variation of the origin positions for each row can be corrected similar to that described in Section \ref{sec:nonlinearstack}.

\begin{figure}[ht!]
	\centering
	\caption{ Illustration of distortions in two frames with mutually perpendicular directions of fast scanning. Rows in the left image are rigid, and their origin can be corrected with regard to the reference in the right image, although the correspondence is not perfect. Similarly, the origin of rigid columns in the right image can be corrected using the reference in the left frame. }
	\includegraphics[width=1.0\textwidth]{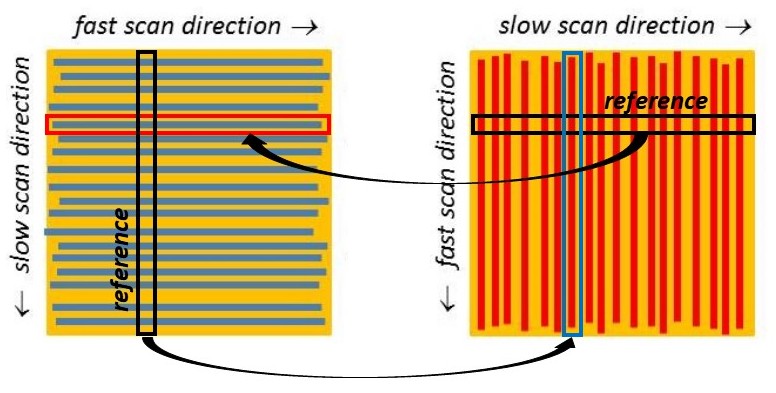}
	\label{fig_12}
\end{figure}
Changing the fast scan direction offers an important benefit: the reference in the companion frame is expected to be much closer to the ground truth than that in Section \ref{sec:nonlinearstack}. Indeed, a rigid row in the left frame is referenced to the slice of rigid columns in the right frame (Fig. \ref{fig_12}).
Although the fine details in such slice might not  coincide exactly with those in the rigid row, their \textit{middle} positions are averaged over many rigid columns with quasi-randomly shifted origins, thus it may serve as a good reference. Clearly, this strategy is only efficient if any systematic variations, like linear drift, are preliminarily extracted from both frames.

\begin{figure}[ht!]
	\caption{  A clear irregularity in the image after linear correction (left) is sealed after correction for non-linear distortions (right). The left image is taken from Fig. \ref{fig_10}, right.}
	\includegraphics[width=0.9\textwidth]{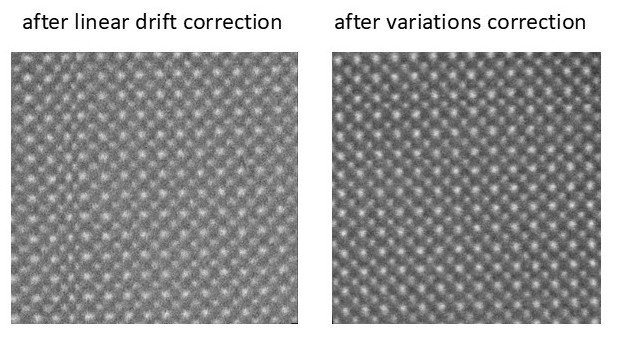}
	\label{fig_13}
\end{figure}

Once the left frame is corrected, it may serve as a reference for aligning rigid columns in the right frame, and the process is repeated until the two images converge. Similar to Section \ref{sec:nonlinearstack}, iterations are stable if the shifts are only partially corrected at each iteration step. The complete restoration of the shifts is only possible in the last iterations, when the images converge.  An example of the reconstructed image is shown in Fig. \ref{fig_13}.

\subsubsection{Frame Resizing}

Similar to Section \ref{sec:align_stack}, the frame size should be enlarged to keep all original pixels or reduced if blanked areas should be avoided:
\begin{align}
	W^L & = \max_{O,R}([\max(x)]_O, [\max(x)]_R) - \max_{O,R}([\min(x)]_O,[\min(x)]_R)) & \\
	H^L & = \max_{O,R}([\max(y)]_O, [\max(y)]_R) - \max_{O,R}([\min(y)]_O,[\min(y)]_R)) & \\
	W^S & = \min_{O,R}([\max(x)]_O, [\max(x)]_R) - \max_{O,R}([\min(x)]_O,[\min(x)]_R)) & \\
	H^S & = \min_{O,R}([\max(y)]_O, [\max(y)]_R) - \max_{O,R}([\min(y)]_O,[\min(y)]_R)) & \\
\end{align}

\section{Correction in Selected Area}

\begin{figure}[ht!]
	\caption{Correction of a crystalline particle placed in a fraction of the acquired field of view. The area selected for correction is outlined in orange. Only rigid rows within the dashed orange region are corrected.}
	\includegraphics[width=0.6\textwidth]{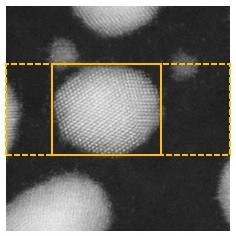}
	\label{fig_16}
\end{figure}

In many cases, correction is needed only within a certain area of the acquired images. This is illustrated  in Fig. \ref{fig_16}
where a crystalline particle  surrounded by an amorphous background is scanned multiple times. The latter area is out of interest, moreover, attempts to align the scanning rows there would likely add undesired artifacts.

Thus, it is worth selecting a region of interest around the particle and correct the image stack within this limited range of scanning rows. The correction proceeds as described above with the only difference that the pause $\Delta T$ between scanned \textit{selected} rectangles  cannot be assumed zero anymore, even if the nominal frames are acquired directly one after another. Instead, it should be set as

\begin{align}
	\Delta T =(1 - \frac{h}{H}) T
\end{align}
where $h$ is the height of the selected area, $H$ is, as usual, the nominal height of the frame, and $T$ is the time for acquisition of one frame.

This modification must be introduced if a linear drift is evaluated from cross-correlation. In cases of drift estimation from FFT, this is irrelevant.

The extension to the case of alternating scan directions is evident.

\section{Summary of Correction Strategies}

Depending on the images sizes, the applied dwell time, the levels of the drift and the magnitude of irregular distortions, different strategies can be used. All of them imply acquisition of several frames followed by reconstruction of a 'true' image.

\begin{enumerate}
	\item If the drift is noticeable, correct each frame for linear drift \textit{via} affine transformation as in Section \ref{sec:linearstack}.
	      The drift value can be retrieved for each frame by cross-correlating with the neighboring frames. Summing up the frames is not recommended.

	      \textit{Variants:}
	      \begin{itemize}
		      \item Assume the constant drift for the whole series. Its value can be retrieved by averaging cross-correlation among all frames.
		      \item Acquire images with different scan rotation and correct for linear drift as in Section \ref{sec:linearalternated}. Acquiring two images might be sufficient in this case, however, images must be nearly square to avoid excessive distortions of the field of view.
		      \item For atomically resolved images, retrieve the linear drift from FFT as in Sections \ref{sec:FFT_stack}, \ref{sec:FFT_rotated}. This is typically more robust against artifacts.
	      \end{itemize}

	\item If the drift is negligible or linearly corrected, but there are evident irregular distortions, apply non-linear correction for irregular variations as in Sections \ref{sec:nonlinearstack} and \ref{sec:nonlinearrotated}. Sum up all frames within the area of no blanked segments.
\end{enumerate}

\section{Applications to Spectrum-Imaging and 4DSTEM}

The above listed strategies can be also applied to STEM EELS or EDX spectrum images \cite{Wang2018} as well as to 4DSTEM with few important remarks.

\begin{itemize}

	\item Spectrum-imaging and 4DSTEM typically takes a longer dwell time, thus all distortions might be more pronounced and require more extensive correction.
	\item Spectrum-imaging is often accompanied by simultaneous acquisition of 2D dark- or bright-field STEM images, which  share the same field of view, dwell time and sampling. These images can be readily used to retrieve STEM distortions and apply corrections to \textit{both} dark/bright field images and the spectral data cube. In case dark/bright field images are not available, a kind of a bright-field image can be produced from the spectrum-image by integrating over all energy range. Similarly, integrating over a part of the diffraction patterns can be used for the correction of 4DSTEM.
	\item Spectrum-images and 4DSTEM are often augmented  by an overview STEM image (sometimes called a Survey Image), which is used to select the area under spectroscopic investigation. This image is always taken with a much smaller dwell time and usually show minimal distortions. Thus, it can be used as a ground truth for the correction, similar to what is described above. Clearly, the reference area in the survey image must be resampled (interpolation needed) in order to fit the pixel size to that in the spectrum-image or 4DSTEM.
	\item In 2D images, the sub-pixel correction delivers better accuracy  and is therefore preferred. However, sub-pixel interpolation breaks the assumption of uncorrelated noise that is often used in  statistical treatment of large datasets. In Appendix, we show that this issue is not restrictive and might affect the results only when a very local area with a small number of pixels is statistically treated.

\end{itemize}

\section*{Acknowledgment}

Pavel Potapov acknowledges the financial support of CEOS GmbH.

\section*{Appendix: Statistical Consequences of Sub-Pixel Correction of Spectrum-Images and 4DSTEM}

Ignoring specific properties of used detectors, the signals acquired in different pixels of STEM image, spectrum-images or 4DSTEM can be assumed statistically independent:

\begin{equation} \label{eqA1}
	s_{ij} = s^*_{ij} +n_{ij}
\end{equation}
where $s^*_{ij}$ is the true signal and $n_{ij}$ is a random noise, mostly of the Poisson nature.

The bilinear interpolation breaks this assumption and makes the noise correlated:
\begin{equation} \label{eqA2}
	s_{ij} = \sum_{k=0}^{4}{c_{kij}s^*_{kij}} +\sum_{k=0}^{4}{c_{kij}n_{kij}}
\end{equation}
where $c_{kij}$ are interpolation coefficients for $ij$-th pixel and $k$ scrolls through the four $(i-1,j-1), (i+1,j-1), (i-1,j+1), (i+1,j+1)$  pairs of the nearest pixel neighbors.

Most fitting procedures and statistical treatments require that the deviations from the model approach zero in terms of quadratic differences while coupling coefficients $c_{kij}$ might create a bias in the noise term $n_{ij}$.

However, as noise correlation is observed within the vicinity of only 4 nearest neighbors, this bias will quickly extinct with the number of considered pixels. As most statistical treatments such as PCA involve the number of pixels $>> 4$, this bias can be neglected.

\bibliographystyle{unsrt}
\bibliography{references}

\end{document}